\newcommand{\beq}{\begin{equation}}
\def\seq{\end{equation}}
\def\beqs{\begin{eqnarray}}
\def\seqs{\end{eqnarray}}

\def\vf{v_F}

\def\D{\Delta}
\newcommand{\DD}{\Delta_0}

\documentclass[aps,prl,twocolumn,superscriptaddress,floatfix]{revtex4}
\usepackage{graphics,epsf}
\begin{document}

\title{Tail states in clean superconductors with magnetic impurities}
\author{A.~V.~Shytov}
\affiliation{Kavli Institute for Theoretical Physics, University
of California, Santa Barbara, CA 93106\\
and L.~D.~Landau Institute for Theoretical Physics, 2 Kosygin St.,
Moscow, Russia 117334}
\author{I.~Vekhter}
\affiliation{Theoretical Division, MS B262, Los Alamos National
Laboratory, Los Alamos, NM 87545}
\author{I.~A.~Gruzberg}
\affiliation{Department of Physics, The University of Chicago,
Chicago, IL 60637}
\author{A.~V.~Balatsky}
\affiliation{Theoretical Division, MS B262, Los Alamos National
Laboratory, Los Alamos, NM 87545}
\begin{abstract}
We analyse the behavior of the density of states in a singlet
$s$-wave superconductor with weak magnetic impurities in the clean
limit. By using the method of optimal fluctuation and treating the
order parameter self-consistently we show that the density of
states is finite everywhere in the superconducting gap, and that
it varies as $\ln N(E)\propto -|E-\DD|^{(7-d)/4}$ near the mean
field gap edge $\DD$ in a $d$-dimensional superconductor. In
contrast to most studied cases the optimal fluctuation is strongly
anisotropic.
\end{abstract}
\date{\today}
\maketitle

One of the most intriguing problems in modern condensed matter
physics is the combined effect of correlations and disorder on the
ground state and the electronic properties of solids. Interplay of
superconductivity and impurity scattering is an example of a
situation where such an effect is relevant experimentally, and
where it has been studied for over forty years, beginning with the
seminal papers by Anderson \cite{Anderson59} and Abrikosov and
Gor'kov (AG) \cite{AG}.
 Despite such a long history,
however, the spectral properties even in the simplest case of a
singlet $s$-wave superconductor continue to attract attention.

Since nonmagnetic impurities do not break the time-reversal
symmetry, the ``hard'' gap, $\DD$, in the single particle
excitation spectrum is unaffected by weak potential scattering
\cite{Anderson59}. At the same time scattering by magnetic
impurities destroys the phase coherence in the superconducting
state, and, consequently, leads to the suppression of the gap, and
of the superconducting order parameter. AG analysed magnetic
scattering using self-consistent Born approximation, and concluded
that a hard gap in the energy spectrum survives up to a critical
(average) concentration of weak magnetic impurities; it is
followed, upon increasing impurity concentration, by a narrow
region of gapless superconductivity, and then by destruction of
the superconducting condensate.

After some early work \cite{LO}, resurgence of interest in this
problem started with the paper of Balatsky and Trugman \cite{BT},
who argued that rare regions, where impurity concentration is
sufficient to locally destroy superconductivity, yield a finite
density of states (DOS) at the Fermi level. Their work was
followed by other analyses \cite{Simons}. Here we elucidate the
nature of the subgap states, and provide a quantitative analysis
of the energy profile of the DOS in $s$-wave superconductors with
weak magnetic impurities.

In the AG theory the effect of disorder is controlled by a
dimensionless parameter, $\Delta\tau_s$, where $\tau_s$ is the
scattering time due to magnetic impurities, and $\Delta$ is the
value of the superconducting order parameter. In particular, the
single particle spectral gap is $\DD=\Delta
(1-(\Delta\tau_s)^{-2/3})^{3/2}$, indicating the onset of the
gapless superconductivity at $\Delta\tau_s=1$. In the regime
studied here, $\Delta\tau_s\gg 1$, the AG theory predicts a gapped
quasiparticle spectrum.

These results are obtained by carrying out a standard impurity
averaging procedure. It is clear, however, that, among all the
realizations of the impurity distribution, there exist regions
where the resulting potential generates localized quasiparticle
states at an arbitrary energy below the gap edge. Such localized
states were extensively studied in doped semiconductors
\cite{Lifshitz,Gredeskul}. For a particular energy, $E$, the most
probable (albeit still rare) configuration of impurities that
creates a state at $E$, and therefore contributes the most to the
DOS, $N(E)$, is called the optimal fluctuation (OF)
\cite{Lifshitz}. Such rare regions provide nonperturbative
corrections to the DOS determined in the framework of
self-consistent Born approximation.

We employ the OF method in a singlet $s$-wave superconductor with
magnetic impurities. We also consider the self-consistent
suppression of the superconducting order parameter; it is known
that in $d$-wave superconductors it significantly affects the
low-energy DOS \cite{Atkinson}. We find that the density of states
is finite everywhere in the superconducting gap. Just below the AG
gap edge, the DOS is $N(E)/N_0\propto
\exp\left(-|E-\DD|^{(7-d)/4}\right)$, where $N_0$ is the normal
state DOS, and $d$ is the number of spatial dimensions. In
contrast to other known cases, the OF is anisotropic, with its
transverse size much smaller than the longitudinal extent.

We consider a singlet $s$-wave superconductor. In the 4-space of
the wave functions $\left(\psi^\star_\uparrow ({\mathbf{r}}),
    \psi^\star_\downarrow ({\mathbf{r}}), \psi_\uparrow ({\mathbf{r}}),
    \psi_\downarrow (\mathbf{r})\right),$
the mean field hamiltonian is
\begin{equation}\label{hamiltonian}
\widehat H=\widehat\xi\tau_3+\Delta({\mathbf{r}})\tau_1\sigma_2
+\widehat U
\end{equation}
Here  $\widehat\xi=-\nabla^2/(2m)-\mu$ is the kinetic energy of a
quasiparticle with respect to the Fermi level, $\mu$, and $\tau_i$
and $\sigma_i$ are the Pauli matrices in the particle-hole and the
spin space respectively, so that $\tau_i \sigma_j$ is a 4$\times
$4 direct product.

Potential due to impurities, $\widehat U$, includes both potential
and spin-flip scattering processes. When the potential scattering
is dominant (motion of quasiparticles in the OF is diffusive)
properties of the low-energy states were explored in Ref.
\cite{Simons}. However, there exists experimental evidence that in
some situations the magnetic scattering is dominant: upon
increasing the impurity concentration the increase in residual
resistivity ratio correlates with the suppression of the
superconducting transition temperature \cite{Edelstein}. Guided by
this insight, we consider only the magnetic scattering, and expect
our results to remain valid for as long as it is stronger than, or
is of the order of, potential scattering. Hence we write $\widehat
U= {\bf U}({\mathbf{r}})\cdot \mathbf{s}$, where $\mathbf{s}$ is
the electron spin operator, ${\bf U}({\mathbf{r}})=\sum_i J {\bf
S}_i\delta({\mathbf{r}} - \mathbf{r}_i)$, $J$ is the exchange
constant, and ${\bf S}_i$ is the localized impurity spin at a site
$i$.

The main physical difference between our analysis and that of Ref.
\cite{Simons} lies in this choice of the scattering potential. In
the regime studied here the mean free path significantly exceeds
the coherence length, and hence the motion of the quasiparticles
in the optimal fluctuation is ballistic, leading to a
substantially different physical picture of the tail states and
the optimal fluctuation, and to a different energy dependence of
the density of states.

The states with energy $E\lesssim \DD$ exist in rare regions where
the amplitude of the impurity potential differs significantly from
its typical value. Therefore in determining $N(E)$ it is
sufficient to consider only such configurations of the impurity
potential, for which $E$ is the lowest quantum mechanical energy
level; fluctuations where $E$ is the energy of a higher bound
state are exponentially less probable \cite{Gredeskul}. Also, in
essentially all the energy range below the gap the size of the
optimal fluctuation is significantly greater than the distance
between impurities, so that the exact impurity potential can be
replaced by a smooth function, averaged over regions containing
many impurities, but smaller than the characteristic size of the
wave function in the optimal fluctuation \cite{Gredeskul}.

Hence we consider an uncorrelated potential with a gaussian probability
density
\begin{equation}\label{gauss}
P[{\mathbf U}]\propto\exp\left(-\frac{1}{2U_0^2}\int d^d{\mathbf r}
{\mathbf U}^2(\mathbf{r})\right),
\end{equation}
where $U_0^2=n_{imp} J^2 S(S+1)/3$ is related to the scattering
time via $\tau_s^{-1}=2\pi N_0 U_0^2$, and $n_{imp}$ is the
average impurity concentration. We ignore interactions between the
magnetic impurities: it was shown in Ref. \cite{Larkin} that the
RKKY interaction and glassy behavior of impurity spins modify the
AG results very weakly. We also do not include quantum dynamics of
the impurity spins, and therefore cannot account for the Kondo
effect. This is justified either when the Kondo temperature of
individual impurity sites is much smaller than the superconducting
transition temperature, $T_K\ll T_c$ (and depletion of states at
the Fermi level prevents screening of the local moment), or in the
opposite limit, $T_K\gg T_c$, when the moments are quenched
already in the normal state \cite{MH}.

The density of localized states in the fluctuation region of the
spectrum is then \cite{Gredeskul}
\begin{equation}
    \label{DOS}
    N(E)=\int {\cal D} {\mathbf U} P[{\mathbf U}]
    \delta(E-{\cal E}[{\mathbf U}]),
\end{equation}
where ${\cal E}[{\mathbf U}]$ is the lowest energy eigenstate in
the realization $\mathbf{U}$ of the impurity potential. For rare
configurations
the integral is evaluated using saddle point approximation to give
$\ln\left[ N(E)/N_0 \right]\approx - {\cal S}[{\mathbf U}_{opt}]$,
where the optimal fluctuation is obtained by minimizing the
functional
\begin{equation}
\label{action}
    {\cal S}[{\mathbf U}]=\frac{1}{2U_0^2}\int d^d{\mathbf r}
{\mathbf U}^2(\mathbf{r}) +
    \lambda \biggl({\cal E }[{\mathbf U}] -E\biggr)
\end{equation}
with respect to both ${\mathbf U}$ and the Lagrange multiplier
$\lambda$. The difficulty in minimizing the action is in the
nonlinear nature of the equations: optimal potential ${\mathbf U}$
depends on the wave function of the particle in this potential.

The method of optimal fluctuation allows for a simple physical
analysis. Consider first a semiconductor. In a potential well of
depth $U$ (all energies are measured from the band edge) and size
$L$ the energy of the localized state is of the order of
$U+1/(mL^2)=E$ ($\hbar=1$). In the optimal fluctuation  $E\sim
U\sim L^{-2}$, so that the action for such fluctuation is ${\cal
S}[U]\approx L^dU^2/U_0^2$, or $\ln\left[ N(E)/N_0 \right]\approx
-|E|^{2-d/2}/U_0^2$. This is exactly the result obtained by
Lifshits, and is confirmed by the solution of the nonlinear
equations for minimization of action in Eq.(\ref{action})
\cite{Lifshitz,Gredeskul}.

The difference between a potential well in a doped semiconductor
and in a superconductor is twofold. First, because of
particle-hole mixing the hamiltonian Eq.(\ref{hamiltonian}) is a
matrix in particle-hole and spin space. Second, we are concerned
with quasiparticles close to the Fermi energy.

We assume that ferromagnetic fluctuation maximizes the effect of
the impurity potential \cite{LT23}. Consequently, we consider such
a fluctuation, and choose the direction of the impurity spins
along the $y$-axis, so that ${\widehat U} =
U(\mathbf{r})\sigma_2$. Performing a spin rotation, $\sigma_2
\rightarrow \sigma_3$ in Eq.(\ref{hamiltonian}), we obtain a
hamiltonian diagonal in the spin space,
\begin{equation}\label{NewHamiltonian}
\widehat H_\pm=
\widehat\xi\tau_3\pm\Delta_0\tau_1 \pm
 U (\mathbf{r}).
\end{equation}
It is therefore sufficient to consider only one spin orientation.
Let us again consider the problem qualitatively, and concentrate
first on the one dimensional case ignoring the suppression of the
order parameter. We linearize the kinetic energy near the Fermi
surface, $\widehat\xi\approx -i v_F (\partial/\partial x)$,  so
that typical kinetic energy in an OF of size $L$ is $\xi\simeq
v_F/L$. Then the energy of a quasiparticle in the optimal
fluctuation (measured from the Fermi energy) is $E\simeq U+
\sqrt{\DD^2 + v_F^2/L^2}$. For the energies close to the
superconducting gap, $(\DD-E)/\DD \ll 1$, the OF is large ($L\gg
\xi_0=v_F/\DD$) and shallow ($|U|/\DD\ll 1$), so that
$E-\DD\approx U+ v_F^2/(\DD L^2)$. Introducing the dimensionless
energy $\epsilon=E/\DD$, we obtain, in analogy with the arguments
above, $|U|/\DD\simeq \xi_0^2/L^2\simeq 1-\epsilon$. Notice that
the size of the fluctuation is indeed $L\simeq
\xi_0/\sqrt{1-\epsilon}\gg \xi_0$. As a result, we find ${\cal
S}[U]\approx LU^2/U_0^2=\DD^2\xi_0(1-\epsilon)^{3/2}/U_0^2$. From
the definition of $U_0$ it follows that
\begin{equation}
    \label{OptimAction1D}
    -\ln\frac{N(E)}{N_0}\approx {\cal S}[U_{opt}]\simeq
    (\DD\tau_s) (1-\epsilon)^{3/2}.
\end{equation}
The energy dependence in Eq.(\ref{OptimAction1D}) is identical to
the result of Lifshits in $d=1$. This follows from the expansion
in $\xi\ll\DD$: even though $\xi\propto 1/L$, the expansion is in
$\xi^2$ .

We now verify these estimates by a considering the energy of a
particle in hamiltonian Eq.(\ref{NewHamiltonian}) for spin ``up''
\begin{equation}\label{NewHamiltonianD}
{\cal E}_+ [U]=\langle \widehat H_+ \rangle=\langle\Psi |
\widehat\xi\tau_3+\Delta_0\tau_1 + U |\Psi \rangle,
\end{equation}
where $\Psi$ is the normalized wave function of the particle.
Minimization of Eq.(\ref{action}) with respect to $U$ gives
\begin{equation}
\label{equation U} U(x)=-\lambda U_0^2 \langle\Psi| \frac{\delta
\widehat H_+}{\delta U}|\Psi \rangle,
\end{equation}
while minimization with respect to $\lambda$ dictates $\widehat
H_+|\Psi\rangle=E|\Psi\rangle$. We first ignore the
self-consistent suppression of the gap, which means $U(x)=-\lambda
U_0^2 (\Psi^\star(x)\Psi (x))$, where $(\Psi^\star\Psi)$ denotes
the scalar product in particle-hole space. Then the Schr\"odinger
equation takes the form
\begin{equation}\label{EqOnPsi}
\Biggl[ -i \vf \frac{\partial}{\partial x}\tau_3 +\DD \tau_1
-\lambda U_0^2 (\Psi^\star\Psi)\Biggr] \Psi = E\Psi.
\end{equation}
This equation is solved by introducing the bilinear forms
$\Psi^\star(x)\tau_i\Psi(x)$, and yields the optimal fluctuation
\begin{equation}\label{OptimalU}
\frac{U(x)}{2\DD}=-\frac{1-\epsilon^2}{\epsilon + \cosh
(2x\sqrt{1-\epsilon^2}/\xi_0)},
\end{equation}
which corresponds to the value of the action
\begin{equation}
\label{OptimalAction1D} {\cal S}[U]=8\pi (\DD\tau_s) \left[
\sqrt{1-\epsilon^2}-\epsilon\arccos\epsilon\right].
\end{equation}
We immediately notice that for $\epsilon\approx 1$ the length
scale of the optimal fluctuation is $\xi_0/\sqrt{1-\epsilon^2}$,
its depth is $U\sim \DD (1-\epsilon^2)$, and the action ${\cal
S}[U_{opt}]\simeq (8\pi/3) (\DD\tau_s) (1-\epsilon^2)^{3/2}$, in
complete agreement with our estimates above.

We now show that the self-consistent suppression of the order
parameter does not appreciably change our result. Self-consistency
is achieved by including the variation of the gap into the
variational derivative $\delta \widehat H_+/\delta U$. We notice
that, at $T=0$, uniform $U$ does not suppress superconductivity.
Consequently, $\D$ depends on the gradient of the potential. For
$1-\epsilon \ll 1$, the potential varies smoothly, so that $dU/dx
\ll \DD/\xi_0$ can be accounted for perturbatively. The leading
local correction to the gap is
\begin{equation}
\label{localD}
\frac{\delta\Delta(x)}{\DD}=-\frac{1}{6}\frac{\xi_0^2}{\DD^2}
\left(\frac{dU}{dx}\right)^2,
\end{equation}
and the correction to the action for $\epsilon\sim 1$ is
$\delta{\cal S}\simeq (128\pi/315)(\DD\tau_s)
(1-\epsilon^2)^{7/2}$. At lower energies the self-consistent
$\Delta(x)$ has to be computed numerically. The results are
presented in Figs. 1 and 2. It is clear that the suppression of
the gap, even for the states with $E\ll\DD$ is incomplete; since
$|U|\leq 2\DD$, the gap remains at a significant fraction of $\DD$
throughout the OF. This justifies the expansion in the bare
energy, $\xi$, in our qualitative analysis. Consequently, the
optimal action computed self-consistently differs at most by 10\%
from that computed assuming a uniform gap, see Fig. 2.
\begin{figure}
    \epsfxsize=0.85\columnwidth \epsfbox{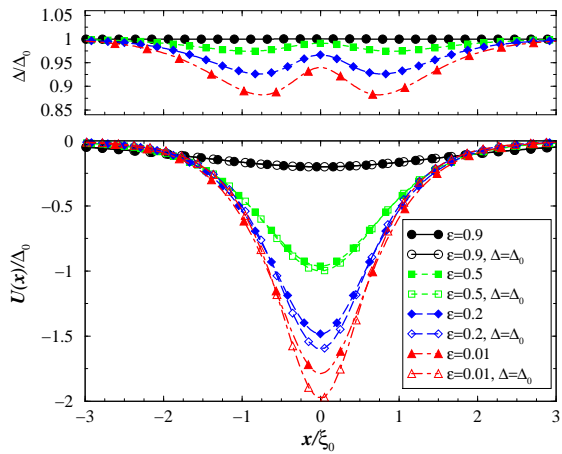}
    \caption{Bottom panel: comparison of the optimal potential for
    $\Delta=\Delta_0$
    from Eq.(\ref{OptimalU})  with that obtained from the self-consistent
    numerical solution, for different
    values of the bound state energy $E=\varepsilon\DD$.
    Top panel: the corresponding self-consistently
    determined profile of the order parameter.
    Maximum in $\Delta(x)$ at $x=0$ corresponds to small
    gap suppression at an extremum of the potential, see Eq.(\ref{localD}).
    }
\end{figure}

Having demonstrated that the qualitative consideration are in
excellent agreement with the full solution of the problem in
$d=1$, we discuss the multidimensional case. In a doped
semiconductor the OF in any $d$ is spherically symmetric
\cite{Lifshitz,Gredeskul}. This results from the balance between
lowering the particle energy in a large and deep fluctuation, and
the probability cost of such an OF.

Since electrons in a superconductor move with the Fermi velocity,
the wave function of the subgap state is concentrated along the
quasiclassical trajectory, which is a chord in a potential of any
shape. Consequently, there is little energy cost in reducing the
size of the OF in the ``transverse'' direction, while the smaller
volume makes such fluctuations more probable. As a result, the
optimal fluctuation is anisotropic, and strongly elongated in one
direction. Choosing this direction as the $x$-axis, we can write
the wave function of the subgap state as $\Psi(x, {\bf
y})=\exp(ik_F x)\Phi(x, {\bf y})$, where ${\bf y}$ denotes the
transverse $d-1$ coordinates, and $\Phi$ is a slowly varying
function. Therefore the kinetic energy of the quasiparticle is
\begin{equation}
    \widehat\xi\Psi \approx -e^{ik_F x}\left( i v_F
\frac{\partial}{\partial x} + \frac{\nabla_y^2}{2m}\right)\Phi
    \sim \left( \frac{v_F}{L_x}+\frac{1}{mL_y^2}\right) \Psi.
\end{equation}
The transverse size of the fluctuation can therefore be reduced
until the second term becomes comparable to the first, i.e.
$L_y\simeq (\lambda_F L_x)^{1/2}$, where $\lambda_F\simeq
k_F^{-1}$ is the Fermi wavelength. Consequently,  we find
$|U|/\DD\sim 1-\epsilon$ and $L_x\sim \xi_0/\sqrt{1-\epsilon}$,
and
\begin{equation}
    \label{dAction}
    {\cal S}[U_{opt}]\simeq L_x L_y^{d-1}\frac{U^2}{U_0^2}\simeq
    (\DD\tau_s)\left(\frac{E_F}{\DD}\right)^{\frac{d-1}{2}}
    \left(1-\epsilon\right)^{\frac{7-d}{4}},
\end{equation}
where $E_F$ is the Fermi energy. Eq.(\ref{dAction}) is the main
result presented here. The action for the anisotropic fluctuation
is smaller than that for an isotropic OF, by a factor of
$(E_F/\DD)^{(d-1)/2} (1-\epsilon)^{-(d-1)/4}$, so that the
corresponding DOS is exponentially higher.

\begin{figure}
    \epsfxsize=0.85\columnwidth \epsfbox{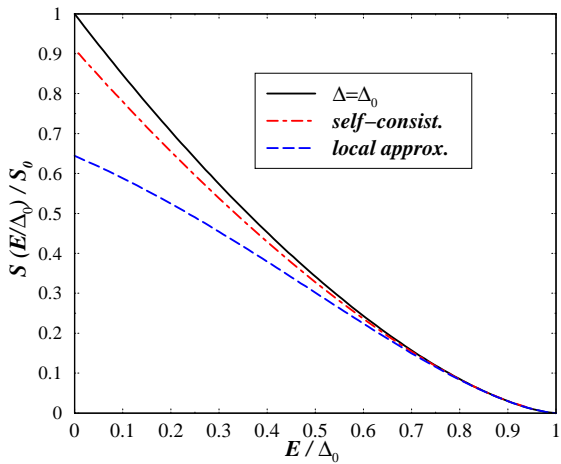}
    \caption{Optimal action, normalized by
    ${\cal S}_0=8\pi (\Delta_0\tau_s)$. Local approximation:
    $\Delta(x)=\Delta_0+\delta\Delta(x)$, with $\delta\Delta(x)$
     from Eq.(\ref{localD}).}
\end{figure}

There are two limitations on the validity of the results obtained
here. First, since the optimal fluctuation is a result of a saddle
point approximation for the functional integral, Eq.(\ref{DOS}),
it is only valid when ${\cal S}[U_{opt}]\gg 1$, or
\begin{equation}
    \label{Valid1}
    1-\epsilon \gg (\DD\tau_s)^{\frac{4}{d-7}}
    \left(\frac{\DD}{E_F}\right)^\frac{2(d-1)}{7-d}.
\end{equation}
For $d=1$ this condition becomes $1-\epsilon\gg
(\DD\tau_s)^{-2/3}$, while for $d=3$ it does not depend on the gap
amplitude, $1-\epsilon\gg (k_F l)^{-1}$, where $l=v_F\tau_s$ is
the mean free path. In $d=1$ the region of validity is extended by
almost an order of magnitude in comparison to this estimate as the
action has a large numerical factor $\approx 24$.

Second, when the characteristic size of the OF $L\geq l$, our
assumption about the ballistic motion in the fluctuation is
invalid, and a crossover to the diffusive regime studied in Ref.
\cite{Simons} occurs for $1-\epsilon \leq (\DD\tau_s)^{-2}$ in any
dimension. When $d=1$ for $(\DD\tau_s)\geq 5$ the saddle point
approximation becomes invalid before the diffusive regime is
reached. For $d>1$ and typical $\DD/E_F\simeq 10^{-3}$, the OF
method works up to the crossover. Taking $(\DD\tau_s)\sim 10$, we
find that our results hold to within 1\% of $\DD$, while the
expansion in $\xi$ is quantitatively valid for $\epsilon\geq 0.9$,
and qualitatively for $\epsilon\geq 0.75$, providing a significant
window of applicability for our DOS.

Experimental verification of our results, and the underlying
physical picture of scattering on randomly distributed impurities,
requires averaging the tunneling conductance over regions
containing many impurities, and is best done at energies just
below $\Delta$. We suggest averaging the tunneling spectra,
obtained from Scanning Tunneling Spectroscopy, over several
distinct areas of the sample, each containing a large number of
impurities.

To summarize, we have analysed the density of subgap states in an
s-wave superconductor with weak magnetic impurities using the
method of the optimal fluctuation. We concentrated on the clean
limit, $l\gg \xi_0$, when the motion of particles in the optimal
potential is ballistic. We find that the optimal fluctuation in
this case is strongly anisotropic, and that the density of states
varies as a stretched exponential below the gap edge, with a power
that depends on the dimension.

This research was supported by the DOE under contract
W-7405-ENG-36, by the NSF Grant PHY-94-07194, and by Pappalardo
Fellowship. We are grateful to the ITP Santa Barbara for
hospitality and support.


\begin{thebibliography}{99}

\bibitem{Anderson59} P.~W.~Anderson, J. Phys. Chem. Solids {\bf
11}, 26 (1959)
\bibitem{AG} A.~A.~Abrikosov and L.~P.~Gor'kov, Sov. Phys. JETP
    {\bf 12}, 1243 (1961).
\bibitem{LO} I.~O.~Kulik and O.~Yu.~Itskovich, Sov. Phys. JETP
{\bf 28}, 102 (1969); A.~I.~Larkin and Yu.~N.~Ovchinnikov, {\it
ibid.} {\bf 34}, 1144 (1971).

\bibitem{BT} A.~V.~Balatsky and S.~A.~Trugman, Phys. Rev. Lett.
{\bf 79}, 3767 (1997)
\bibitem{Simons} A. Lamacraft and B.~D.~Simons, Phys. Rev. Lett.
    {\bf 85}, 4783 (2000); Phys. Rev. B {\bf 64}, 014514 (2001);
    I.~S.~Beloborodov, B.~N.~Narozhny, and I.~L.~Aleiner, Phys.
    Rev. Lett. {\bf 85}, 816 (2000).

\bibitem{Lifshitz} I.~M.~Lifshitz, Sov. Phys. Usp. {\bf 7}, 549 (1965);
    Sov. Phys. JETP {\bf 26}, 462 (1968); J.~Zittarz and J.~S.~Langer,
    Phys. Rev. {\bf 148}, 741 (1966); B.~I.~Halperin and M.~Lax,
    {\it ibid.} 722.

\bibitem{Gredeskul} I.~M.~Lifshits, S.~A.~Gredeskul, and
        L.~A.~Pastur, \textit{Introduction to the theory of
        disordered systems}, (John Wiley \& Sons, New York, 1988).


\bibitem{Atkinson} W.~A.~Atkinson, P.~J.~Hirschfeld, and
A.~H.~MacDonald, Phys. Rev. Lett. {\bf 85}, 3922 (2000).

\bibitem{Edelstein} M.~A.~Woolf and F.~Reif, Phys. Rev. {\bf 137},
    A557 (1965); A.~S.~Edelstein, Phys. Rev. Lett. {\bf 19}, 1184
    (1967).

\bibitem{Larkin} A.~I.~Larkin, V.~I.~Mel'nikov, and D.~E.~Khmelnitskii,
    Sov. Phys. JETP {\bf 33}, 458 (1971); V.~Galitski and
    A.~I.~Larkin, cond-mat/0204189.

\bibitem{MH} E.~M\"uller-Hartmann and J.~Zittartz, Phys. Rev. Lett.
{\bf 26}, 428 (1971).

\bibitem{LT23} We conjecture that the ferromagnetic fluctuation is the most
advantageous of spin ordered OFs. It gives a higher DOS than the
paramagnetic OF, see I.~Vekhter, A.~V.~Shytov, I.~A.~Gruzberg, and
A.~V.~Balatsky, cond-mat/0210421, to appear in Physica B.
\end{thebibliography}
\end{document}